\documentclass[preprint,floatfix]{revtex4}
\usepackage{graphicx}
\usepackage{epsf}

\newcommand{\e}{{\rm e}}

\newcommand{\dte}{{\epsilon}}

\newcommand{\x}{{\bf x}}

\newcommand{\br}{{\bf r}}

\newcommand{\bea}{\begin{eqnarray}}
\newcommand{\eea}{\end{eqnarray}}
\newcommand{\be}{\begin{equation}}
\newcommand{\ee}{\end{equation}}
\newcommand{\ba}{\begin{eqnarray}}
\newcommand{\ea}{\end{eqnarray}}

\newcommand{\nn}{\nonumber}
\newcommand{\la}{\label}

\def\t1{e_{_T}}
\def\v1{e_{_V}}

\begin{document}
%\tightenlines
\title{A topological proof that there is no sign problem in one dimensional
Path Integral Monte Carlo simulation of fermions}

\author{Siu A. Chin}

\affiliation{Department of Physics and Astronomy,
Texas A\&M University, College Station, TX 77843, USA}

%\date{\today}
\begin{abstract}

This work shows that, in one dimension, due to its topology,
a closed-loop product of short-time propagators is always positive,
despite the fact that each anti-symmetric free fermion 
propagator can be of either sign.

\end{abstract}
\maketitle

\section {Introductions}

In their first path-integral Monte Carlo (PIMC) simulation of fermions
in a one-dimensional harmonic oscillator,  
Takahashi and Imada
were surprised that, ``In the calculation of one-dimensional
fermions, we do not find any cases of negative
weight function in ten thousands Monte Carlo steps
even at low temperature"\cite{tak84}. They surmised that
their situation maybe similar to the one-dimensional 
lattice fermions studies of Hirsch, Scalapino, Sugar and Blankenbecker\cite{hir81}.
The two are not similar. By a clever lattice arrangement, the matrix elements of 
Hirsch {\it et al.}'s lattice fermions can be chosen to be 
positive\cite{hir81}, while the 
free fermion propagator used by Takahashi and Imada can have either sign.
However, 24 years earlier, Girardeau\cite{gir60} has shown that, in
one dimension, the ground state wave function of $N$ impenetrable bosons 
$\{x_1,x_2 \cdots x_n\}$, which vanishes whenever  $x_i=x_j$,
is the same as the modulus of the
ground state wave function of $N$ {\it free} fermions:
\be
\psi_0^B=|\psi_0^F|.
\ee 
This means that, in one dimension, $N$ {\it interacting} fermions
can always be mapped into the ordered subspace
\be 
x_1<x_2<\cdots<x_N,
\la{sub}
\ee
with vanishing wave function at $x_i=x_{i+1}$.
The ground state wave function can then be taken to be positive, 
the same as that of $N$ impenetrable, interacting bosons\cite{neg88}.
Alternatively, one can view the subspace (\ref{sub}) 
as having the correct wave function nodes at $x_i=x_{i+1}$,
thereby reduced a many-fermion problem, 
to that of a many-boson problem in a single nodal region\cite{cep91}. 
Both views explain that fermions in one dimension do not have the sign problem
because it is basically a boson problem.

However, these two views do not explain why there is no
sign problem {\it specifically} for PIMC simulations,
despite the fact that the anti-symmetric free fermion propagator can have either
sign and that the simulation is not restricted to any particular nodal region. 

This work found that there is a surprisingly simple, but overlooked topological proof,
that there is no sign problem for PIMC simulation of one dimensional fermions. 
This topological explanation
is related to the original insight of Girardeau\cite{gir65}, 
that any statistics is permissible in one dimension, but only Fermi-Dirac or Bose-Einstein
statistics is mandated in more than one dimension.

\section {Fermion Path Integral Monte Carlo}

Consider the single particle imaginary time Schr\"odinger equation
in one-dimension,
\be
-\frac{\partial\psi(x,\tau)}{\partial \tau}=(\hat T+\hat V)\psi(x,\tau)=
\left(-\frac12\frac{\partial^2}{\partial x^2}
+V(x)\right)\psi(x,\tau),
\ee
with dimensionless spatial variable $x$ and imaginary time $\tau$.
In PIMC, one is
interested in extracting the ground state wave function squared $\psi^2_0(x)$ 
and energy $E_0$ from the {\it diagonal} element of the 
imaginary time propagator at the large time limit:
\be
\lim_{\tau\rightarrow\infty}G(x,x;\tau)\longrightarrow\psi_0^2(x)\e^{-\tau E_0}+ \cdots,
\ee
where
\be
G(x^\prime,x;\tau)=\langle x^\prime|\e^{-\tau(\hat T+\hat V)}|x\rangle
=\sum_n\psi_n^*(x^\prime)\psi_n(x)\e^{-\tau E_n}.
\ee
Since $G(x^\prime,x;\tau)$ is generally unknown,  
it is approximated by $k$ short-time propagators via
\ba
G_{k}(x^\prime,x;\tau)&=&\langle x^\prime|(\e^{-\dte(\hat T+\hat V)})^k|x\rangle\nn\\
&=&\int_{-\infty}^{\infty} dx_1\cdots dx_{k-1}\,
G_1(x^\prime,x_1;\dte)G_1(x_1,x_2;\dte)\cdots G_1(x_{k-1},x;\dte),
\la{mb}
\ea
where $\dte=\tau/k$ and $G_1(x',x,\dte)$ is 
usually the second-order short-time approximation of 
$\langle x'|\e^{-\dte(\hat T+\hat V)}|x\rangle$,
the {\it primitive approximation} (PA) propagator:
\ba
&&G_1(x',x;\dte)=\langle x'|\e^{-(\dte/2)\hat V}\e^{-\dte\hat T}\e^{-(\dte/2)\hat V}|x\rangle\nn\\
&&\qquad =\frac1{\sqrt{2\pi \dte}}\e^{-(\dte/2) V(x')}
\e^{-(x'-x)^2/(2\dte)}
\e^{-(\dte/2) V(x)}.
\la{pa}
\ea
To generalize the above to $N$ fermions, one replaces
$x$ by $\x=(x_1,x_2 \cdots x_N)$ and $G_1(x',x,\dte)$ by
\ba
&&G_1(\x',\x;\dte)=
\e^{-(\dte/2) V(\x')}
G_0(\x',\x;\dte)
\e^{-(\dte/2) V(\x)},
\la{pap}
\ea
where $G_0(\x',\x;\dte)$ is the anti-symmetric free-fermion propagator
\ba
G_0(\x^\prime,\x;\dte)&=&\frac1{N!}{\rm det}\left(\frac1{\sqrt{2\pi\dte}}
\exp\left[-\frac1{2\dte}(x_i^\prime-x_j)^2\right] \right).
\la{free}
\ea
Note that any pair exchange $x'_i\leftrightarrow x'_j$ ($x_i\leftrightarrow x_j$) interchanges two rows (columns)
of the determinant and hence the sign of $G_0(\x^\prime,\x;\dte)$, while $G_0(\x^\prime,\x;\dte)=G_0(\x,\x^\prime;\dte)$.

\section {No sign problem in one dimension}

The sign of the integrand in the discrete path integral (\ref{mb})
depends only on the product of $k$ free-fermion propagators:
\be
G_0(\x,\x_1;\dte)G_0(\x_1,\x_2;\dte)\cdots G_0(\x_{k-1},\x;\dte).
\la{prodm}
\ee
For extracting $\psi_0^2(\x)$, the propagators must start at $\x$ 
and loop back to $\x$. The integral is that of a closed-end path-integral. 
Consider first, the case of two (spinless) fermions. 
The anti-symmetric free propagator is then
\ba
G_0(x_1^\prime,x_2^\prime,x_1,x_2;\dte)
&=&\frac12\frac1{2\pi\dte}
\det\left(\begin{array}{cc}
	\e^{-\frac1{2\dte}(x'_1-x_1)^2} &  \e^{-\frac1{2\dte}(x'_1-x_2)^2}\\
	\e^{-\frac1{2\dte}(x'_2-x_1)^2} & \e^{-\frac1{2\dte}(x'_2-x_2)^2}
\end{array}\right)\nn\\
&=&\frac12\frac1{2\pi\dte}\e^{-\frac1{2\dte}\left[(x_1^\prime-x_1)^2
	+(x_2^\prime-x_2)^2\right] }
\left(1
-\e^{-\frac1{\dte}(x_1^\prime-x^\prime_2)(x_1-x_2) } \right).
\la{prop2}
\ea
Thus $G_0(x_1^\prime,x_2^\prime,x_1,x_2;\dte)< 0$
if and only if
\be
(x_1^\prime-x_2^\prime) (x_1 - x_2)< 0,
\la{node}
\ee
{\it i.e.}, either
$x_1^\prime>x_2^\prime$ and $x_1<x_2$ or vice versa. 
This means that the prime and unprime positions are on opposite sides of 
the line $x_1=x_2$ dividing the $x_1-x_2$ plane. 

The key contribution of this work is to rephrase the above condition in
topological terms: the propagator is negative when the line connecting the prime and unprime
position of the propagator crosses the line $x_1=x_2$. 
This is shown in part A of Fig.1. For two propagators
$G_0(\x,\x_1;\dte)G_0(\x_1,\x;\dte)$, the line is either not crossed or crossed twice 
and the product is always positive.
The same is true for the product of three propagators $G_0(\x,\x_1;\dte)G_0(\x_1,\x_2;\dte)G(\x_2,\x;\dte)$,
as shown in part B. More generally, any closed-loop product of 
propagators must be positive, as shown in part C,
since topologically, any planar closed curve must intersect an infinite straight line even number of times.

%%%%%%%%%%%%%%%%%%%%%%%%%%%%%%%%%%%%%%%%%%%%%%%%%%%%%%%%%%%%%%%%%%%%%%%%%%%%%%
%\newpage
\begin{figure}
	%	\vspace{0.5truein}
	\centerline{\includegraphics[width=0.8\linewidth]{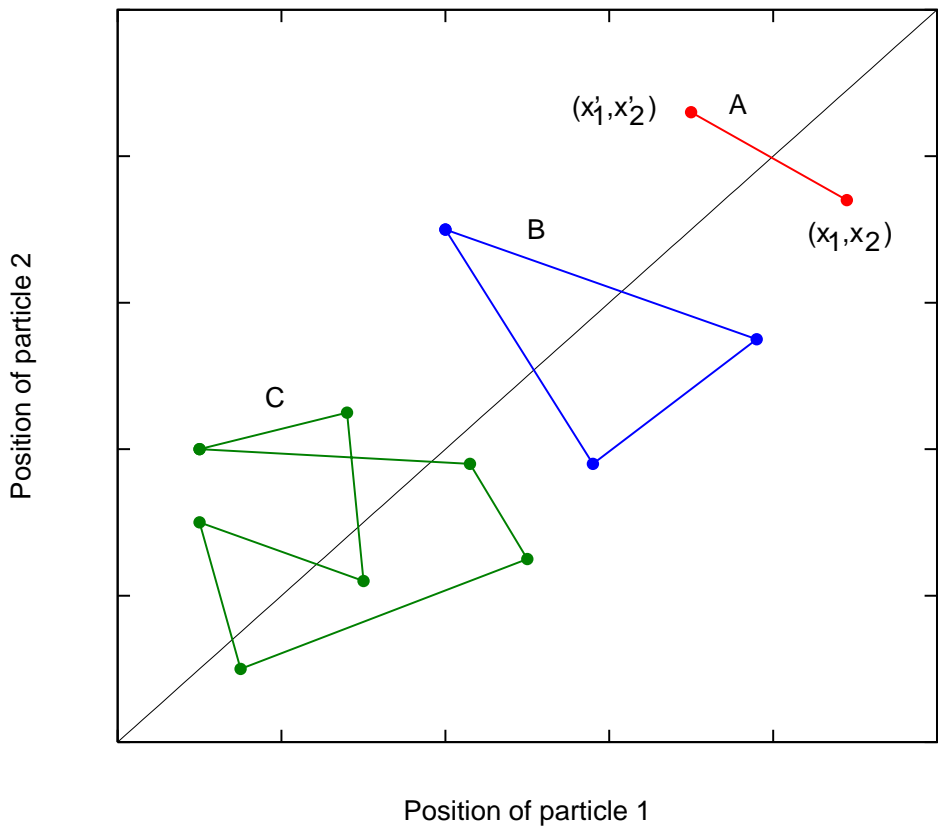}}
	%	\vspace{0.5truein}
	\caption{
		In one dimension, the product of two-particle anti-symmetric propagators in a closed-loop 
		must cross the line $x_2=x_1$ an even number of times.
		\label{oned}}
\end{figure}
%%%%%%%%%%%%%%%%%%%%%%%%%%%%%%%%%%%%%%%%%%%%%%%%%%%%%%%%%%%%%%%%%%%%%%%%%%%%%%

For $N$ fermions, the positions of the anti-symmetric propagator are defined in a $N$-dimensional manifold.
The propagator changes sign whenever its initial and final position cross any one of the
$N(N-1)/2$, ($N-1$)-dimensional 
hyper-planes defined by $x_i=x_j$. Since each such ($N-1$)-dimensional hyper-planes completely divides the 
$N$-dimensional manifold into two halves, any closed curve in the $N$-dimensional manifold must pierce 
each such hyper-plane even number of times. Thus a closed-loop product of free-fermion propagators
for $N$ fermions is also always positive. 

\section {Sign problem in more than one dimension}

In $d$-dimension, one replaces $x_i$ by $d$-dimensional vectors $\br_i=(x_i,y_i,z_i, \cdots)$ and set
$\x=(\br_1,\br_2 \cdots \br_N)$. In this case the anti-symmetric two-fermion free propagator is
\ba
G_0(\br_1^\prime,\br_2^\prime,\br_1,\br_2;\dte)
&=&\frac12\frac1{(2\pi\dte)^d}\e^{-\frac1{2\dte}\left[(\br_1^\prime-\br_1)^2
	+(\br_2^\prime-\br_2)^2\right] }
\left(1
-\e^{-\frac1{\dte}(\br_1^\prime-\br^\prime_2)\cdot(\br_1-\br_2) } \right),
\la{propr2}
\ea
and vanishes whenever\cite{cep91}
\be
(\br_1^\prime-\br^\prime_2)\cdot(\br_1-\br_2)=0.
\la{pinc}
\ee
In two-dimension, the two-fermion propagator is defined in the four-dimensional manifold $(x_1,y_1,x_2,y_2)$,
and vanishes at the {\it coincident} plane\cite{cep91} given by $x_1=x_2$ and $y_1=y_2$. This is
the direct generalization of the one dimensional case. However, in this case, the
coincident plane is only two dimensional, two dimensions less than the full manifold and
therefore does not divide the four-dimensional manifold into disjoint regions\cite{gir65}. 
(This is similar to the case of a line, 
which is two dimensions less than, and therefore cannot divide, the three dimensional Euclidean space.) 
(Away from the coincident plane, the propagator, according to (\ref{pinc}), can also vanishes if the 
relative vector $\br'_{12}=\br_1^\prime-\br^\prime_2$ is perpendicular to the relative vector $\br_{12}=\br_1-\br_2$.
In two dimension, $\br_{12}$ can be oriented at an arbitrary angle $\phi$. Thus away from the coincident plane,
the propagator can additionally vanishes at two one-dimensional circles. This measure zero effect can be ignored.)
Therefore, in the four-dimensional manifold $(x_1,y_1,x_2,y_2)$, 
a closed curve can either pierces the coincident plane, or goes around it.
Thus a closed-loop product of anti-symmetric propagators can be of either sign
and one has a sign problem.

Generalizing this to $N$ particles in $d$-dimension, the propagator is defined in a $Nd$-dimensional manifold.
Any coincident plane is of dimension ($Nd-d$) and cannot fully divide the
$Nd$-dimensional manifold {\it except} for $d=1$. Therefore, the sign problem is generally 
pervasive except in one dimension.

\section {Concluding remarks}

The observation that a $Nd$-dimensional manifold remains connected, despite the existence of
($Nd-d$) dimensional coincident hyper-planes, was Girardeau's\cite{gir65} insight that the conventional proof
for Fermi-Dirac or Bose-Einstein statistics only applies to $d>1$. (The loop-hole for anyon statistics 
in $d=2$ was a later development\cite{lei77,wil82}.) For $d=1$, since each coincident plane completely divides the
manifold, statistics based any permutation symmetry is permissible\cite{gir65}. Here, it provided a simple
proof that there is no sign problem in PIMC simulations of fermions in one dimension.

%%%%%%%%%%%%%%%%%%%%%%%%%%%%%%%%%%%%%%%%%%%%%%%%%%%%%%%
\end{document}